%% file: main.tex
\documentclass[journal]{IEEEtran}

\usepackage{xcolor,soul,framed} 
\usepackage{booktabs}
\colorlet{shadecolor}{yellow}
\usepackage[pdftex]{graphicx}
\graphicspath{{../pdf/}{../jpeg/}}
\DeclareGraphicsExtensions{.pdf,.jpeg,.png}

\usepackage[cmex10]{amsmath}
\usepackage{amssymb}
\usepackage{bm}
\usepackage{array}
\usepackage{mdwmath}
\usepackage{mdwtab}
\usepackage{eqparbox}
\usepackage{url}
\usepackage{cite}
\usepackage{comment}
\usepackage[font=small]{caption}
\usepackage{tikz}
\usepackage{graphicx}
\usetikzlibrary{positioning}
\usepackage[labelformat=simple]{subcaption}
\usepackage{float}

\usepackage{siunitx}
\usepackage[acronym,nonumberlist,toc]{glossaries} 
\usepackage[acronym,style=super,nogroupskip,nonumberlist]{glossaries-extra} 

\input{Acronyms.tex}

\begin{document}

\bstctlcite{IEEEexample:BSTcontrol}
\title{Insights on the Uplink Operation of a 1-bit Radio-Over-Fiber Architecture in Multi-User D-MIMO Communication}
\author{Lise Aabel,
	Giuseppe Durisi,~\IEEEmembership{Senior Member,~IEEE,}
	Frida Olofsson,\\
    Erik B\"orjeson,
	Mikael Coldrey,
	and~Christian~Fager,~\IEEEmembership{Fellow,~IEEE}

	\thanks{The work of L. Aabel, C. Fager, and G. Durisi was supported in part by the Swedish Foundation for Strategic
		Research under Grant ID19-0036 and in part by the Sweden’s Innovation Agency under Grant 2024-02404.}
	\thanks{L. Aabel is with Ericsson AB, 41756 Gothenburg, Sweden, and also with Chalmers University of Technology, 41296 Gothenburg, Sweden
		(e-mail: lise.aabel@ericsson.com).}
	\thanks{M. Coldrey is with Ericsson AB, 41756 Gothenburg, Sweden  (e-mail: mikael.coldrey@ericsson.com).}
	\thanks{F. Olofsson, G. Durisi, E. B\"orjeson, and C. Fager are with Chalmers University of Technology, 41296 Gothenburg,
		Sweden (e-mail: frida.olofsson@chalmers.se;  durisi@chalmers.se; erikbor@chalmers.se; christian.fager@chalmers.se).}
}

\maketitle

\begin{abstract}
	We consider a distributed multiple-input multiple-output (D-MIMO) testbed in which, to enable
	coherent-phase transmission without over-the-air synchronization, the remote radio heads (RRHs)
	are connected to a central unit via a $1$-bit radio-over-fiber fronthaul.
	Specifically, $1$-bit samples of the radio-frequency signal are exchanged over the fronthaul.
	We investigate via both measurements and simulations based on an accurate model of the
	testbed hardware, the capability of the proposed architecture to provide uniform quality
	of services over the coverage area---one of the promises of D-MIMO.
    Our results are encouraging: for the case in which two user equipments (UEs) communicate over the
	same \SI{75}{MHz} signal bandwidth, the measured error-vector magnitude  meets
	the 3GPP New Radio specification of $\mathbf{12.5\%}$ for $\mathbf{16}$QAM across all tested D-MIMO scenarios.
	We also determine that uplink transmission is a potential bottleneck, due to the
	limited dynamic range of the automatic gain controller, which prevents the $1$-bit
	quantizer to benefit from dithering.
	We show that this issue can be mitigated via UE power control.
\end{abstract}



%


\section{Introduction} 
\IEEEPARstart{D}{istributed} \gls{mimo} is a wireless network architecture in which
multiple, spatially distributed \glspl{rrh} serve \glspl{ue} cooperatively. The benefits of
D-\gls{mimo}, compared to conventional centralized architectures, lie in its potential to
increase energy efficiency, reliability, and coverage, as well as in its suitability
to support integrated communication and sensing operations~\cite{bjornson_cellfree,hexaX_23,guo25-02a}.
D-\gls{mimo} architectures can involve different degrees of cooperation between the
\glspl{rrh}~\cite{bjornson_cellfree,guo25-02a}.
However, the full benefits of D-\gls{mimo} can be released only when the distributed
\glspl{rrh}
transmit phase coherent signals to the \glspl{ue}.

Implementing D-\gls{mimo} with phase-coherent transmission requires maintaining \gls{rf}
phase synchronization among all cooperating \glspl{rrh}. In conventional \glspl{rrh},
\glspl{dac} and \glspl{adc} operate on \gls{bb} signals, and frequency conversion from
\gls{bb} to \gls{rf} is performed locally at the \glspl{rrh} using a mixer and a \gls{lo}.
Unfortunately, \glspl{lo} are subject to frequency instabilities and phase noise.
This yields a time-varying drift of the \glspl{lo} phases at the \glspl{rrh}, which needs
to be compensated for~\cite{Nanzer24,larsson24-01a}.
This synchronization problem has been studied to a significant extent in the recent literature.
We provide next a brief review, which focuses on practical implementations of coherent
transmission solutions within the context of \gls{tdd} D-\gls{mimo} systems.

\subsection{Synchronization Methods to Enable Coherent Transmission}
A prominent line of work considers the conventional \gls{rrh} design just described, and
relies on over-the-air synchronization to provide phase-coherent
transmission~\cite{larsson24-01a,Nanzer24,Hamed_2016}.
For example, the architecture presented in~\cite{Hamed_2016} achieves this through a 802.11-inspired
leader-follower approach, in which the follower \glspl{rrh} continuously estimate and compensate
for the phase drift of their \glspl{lo}.
However, as discussed in, e.g.,~\cite{larsson24-01a}, over-the-air synchronization faces
unavoidable scalability issues as the number of \glspl{rrh} grows, for certain topologies.

\begin{table*}[t]
	\caption{TDD multi-user D-MIMO testbeds relying on a wired fronthaul}
	\centering
	\begin{tabular}{ @{}c| c c c c c c @{}}
		\toprule
		Ref.            & Fronthaul type      & RRHs & UEs & Deployment                      & Carrier (GHz) & Bandwidth (MHz) \\
		\midrule
		\cite{Torfs_24} & 1-bit BB-over-fiber & 4    & 2   & $\qty{1}{m}\times\qty{2}{m}$    & 3.68          & 46              \\

		\cite{nec24}    & IF-over-coaxial     & 8    & 8   & $\qty{8}{m}\times\qty{8}{m}$    & 28.25         & 100             \\

		\cite{aabel24}  & 1-bit RF-over-fiber & 3    & 2   & $\qty{1}{m}\times\qty{2}{m}$    & 2.35          & 5               \\

		This work       & 1-bit RF-over-fiber & 6    & 2   & $\qty{3.5}{m}\times \qty{4}{m}$ & 2.35          & 75              \\
		\bottomrule
	\end{tabular}
	\label{tab:prevwork}
\end{table*}

A drastically different approach is considered in~\cite{Torfs_24,nec24,aabel24}, where the
authors focus on less conventional \gls{rrh} designs and leverage the presence of a wired
fronthaul connecting the \glspl{rrh} to a \gls{cu}, to avoid over-the-air synchronization
altogether.

In the testbed described in \cite{Torfs_24},  the \glspl{rrh} are connected to the
\gls{cu} via a fiber-optical fronthaul, over which sigma-delta modulated \gls{bb} signals
are exchanged both in the uplink and the downlink.
Specifically, in the downlink, \gls{bb}
signals are sigma-delta modulated at the \gls{cu} and transmitted over the fronthaul to
the \glspl{rrh}, which perform digital up-conversion using a clock recovered from the data
stream. In the uplink, the received \gls{rf} signals are down-converted in the analog
domain using the clock recovered in the downlink, sampled by an \gls{adc}, and then
sigma-delta modulated and transmitted to the \gls{cu}.
Although this strategy naturally yields synchronization among the \glspl{rrh}, it
results in a complex \gls{rrh} architecture.

In \cite{nec24}, a mm-wave D-\gls{mimo} testbed is presented, in which \glspl{rrh} and
\gls{cu} are connected via coaxial cables over which \gls{if} signals are exchanged, and
through which a
clock signal is provided to the \glspl{rrh}. Each \gls{rrh} then performs up- and
down-conversion locally, using the clock signal provided by the \gls{cu}. Although this
ensures phase coherence, the resulting architecture suffers from
scalability, due to the short distances over which signals can be exchanged over
coaxial cables.
Indeed, the coaxial cables are effective only over distances in the order of \SI{10}{m}.
In contrast, a fiber-optic fronthaul can be deployed over several kilometers.

\subsection{The 1-Bit Radio-Over-Fiber Architecture}
In \cite{aabel24}, we introduced a D-\gls{mimo} network architecture upon which this paper
expands, which we shall hereon refer to as \emph{$1$-bit radio-over-fiber architecture}.
This architecture involves a \gls{cu} that is responsible for performing digital frequency
up- and down-conversion.
Coherent transmission is achieved in this architecture by using only a
central \gls{lo} at the \gls{cu} and by exchanging $1$-bit quantized \gls{rf} signals between the \gls{cu} and the
\glspl{rrh} via an optical fronthaul, which enables the use of on-off intensity
modulation.
A two-level optical signal is generated in the downlink through the use of \gls{rf} sigma-delta modulation, and
in the uplink via \gls{rf} $1$-bit quantization with dithering at the \glspl{rrh}.
While the D-\gls{mimo} architectures described in \cite{Hamed_2016,nec24,Torfs_24} require
delicate clock management---either through optical fiber transmission, coaxial cable
connections, or over-the-air synchronization---our architecture circumvents this issue
entirely. As an additional advantage, the design of the \glspl{rrh} is significantly
simplified, which is beneficial from a scalability perspective.

\subsubsection{Previous Contributions and their Limitations}
The uplink and downlink hardware functionalities of this $1$-bit radio-over-fiber
architecture have been demonstrated through point-to-point measurements in
\cite{aabel24,OlofssonWDM25,IbraMe21,Sezgin2019MIMOtestbed}. Specifically, in
\cite{Sezgin2019MIMOtestbed,OlofssonWDM25} it is demonstrated that satisfactory \gls{evm}
performance can be achieved in the downlink using \gls{rf} sigma-delta modulation, both
using a parallel and a serial optical fronthaul configuration. In~\cite{aabel24,IbraMe21},
it is shown that satisfactory \gls{evm} performance can be achieved also in the uplink,
provided that dithering and oversampling are performed at the~\gls{rrh}.
Dithering involves adding to the received signal a suitably designed signal, whose
frequency and power are optimized, so as to whiten the quantization noise.

On the negative side, it was also shown in~\cite{aabel24} via point-to-point uplink measurements
	that it is not always possible to operate at the optimal signal-to-dither ratio, because of
	the limited dynamic range of the \gls{agc} at the \gls{rrh}.
	As a result, whenever the power of the received signal is outside the dynamic range of the
	\gls{agc}, the uplink \gls{evm} performance deteriorates significantly.

	The multi-user performance of the 1-bit radio-over-fiber architecture is largely
	unexplored.
	In~\cite{aabel24}, because of limitations in the testbed architecture (see Section~\ref{sec:key_improvements}), multi-user functionalities  were demonstrated
	only for a small-scale
	deployment scenario involving a coverage area of $\qty{1}{m}\times \qty{2}{m}$ in which $3$ \glspl{rrh} serve $2$ \glspl{ue}.
	Specifically, we showed the feasibility of reciprocity-based coherent downlink
	transmission over a small bandwidth (\SI{5}{MHz}).
	Hence, it remains unclear whether the limited dynamic range of the \gls{agc} has a
	negative impact on multi-user performance in more realistic deployment scenarios.
	Indeed, in the scenario considered in~\cite{aabel24}, the two \glspl{ue} are received at
	a similar power at all \glspl{rrh} because of their proximity, which makes it possible
	to guarantee a close-to-optimal signal-to-dither ratio for both \glspl{ue} signals at
	each \gls{rrh}.
	Although a hardware model for the \gls{rrh} is presented in~\cite{aabel24}, this model
	does not entirely agree with the measurement results and does not capture the
	limitation in the dynamic range of the \gls{agc}. This prevents a simulation-based analysis
	of its impact on multiuser performance.

\subsubsection{Novel Contributions}
In this article, leveraging targeted improvements on the testbed architecture described
in~\cite{aabel24}, we investigate the multi-user performance of the 1-bit
radio-over-fiber architecture over a wider
bandwidth and within a larger deployment area (see Table~\ref{tab:prevwork} for a comparison
between the testbed considered in the present paper and the wired-fronthaul testbeds
presented in~\cite{Torfs_24,nec24,aabel24}, in terms of fronthaul type, number of
\glspl{rrh}, number of \glspl{ue}, size of the deployment area, carrier frequency, and
bandwidth).
Our specific contributions are as follows:
    \begin{itemize}
		\item
		      We first show that, in the single-\gls{ue} case, potential performance
		      losses caused by the limited dynamic range of the \gls{agc} can be easily
		      avoided, if the \gls{rrh} are appropriately distributed across the
		      coverage area, and the \gls{ue} transmit power is properly selected.
		      Under these conditions, satisfactory uplink \gls{evm}
		      performance can be achieved uniformly over the entire coverage area.
		\item Through extensive measurements involving two \glspl{ue}, placed at difference
		      inter-\gls{ue} distance within the coverage area, we show that satisfactory
		      \gls{evm} uplink performance can also be achieved in the multi-user case. This
		      indicates that the limited dynamic range of the \gls{agc} is not a limiting factor
		      for supporting multi-user operation.
		\item Finally we present an accurate \gls{rrh} hardware model that effectively captures
		      most hardware impairments and yields \gls{evm} predictions closely matching
		      the measured ones.
		      Using this model, we perform simulations to demonstrate how \gls{ue} power
		      control can be employed to ensure a uniform quality of service.
	\end{itemize}

The rest of this article is organized as follows.
In Section~\ref{sec:testbed}, we describe our improved D-\gls{mimo} testbed.
Single-user measurements are described in Section~\ref{sec:su-dmimo}, whereas multi-user
measurements are provided in Section~\ref{sec:mu-dmimo}. The model of the testbed,
together with \gls{ue} power-control simulations based on this model, are described in
Section~\ref{sec:tb_model}.
Finally, we provide some concluding remarks in Section~\ref{sec:concl}.

\begin{figure*}
	\centering
	\includegraphics[width=\linewidth]{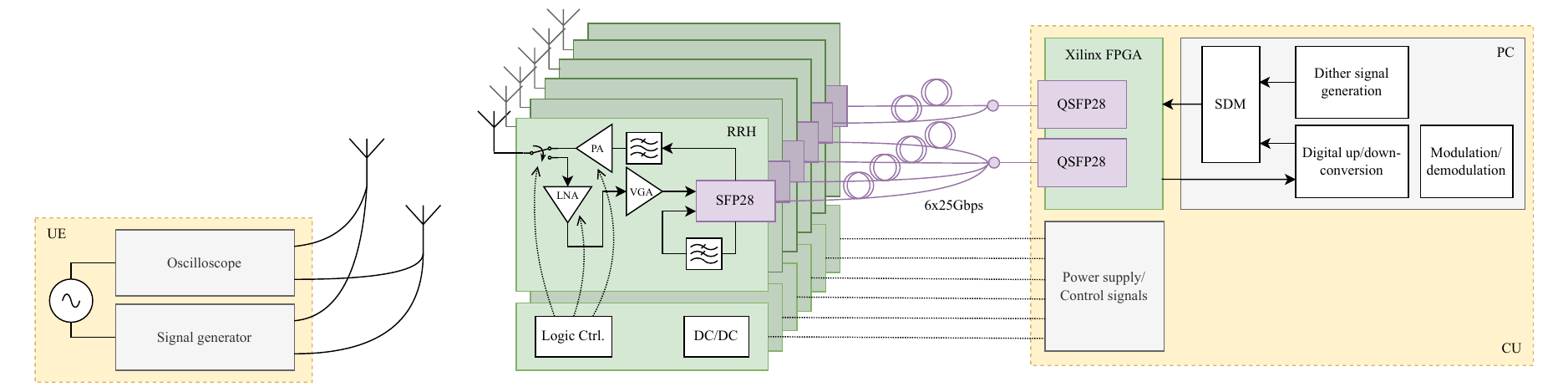}
	\caption{Block diagram describing the improved testbed architecture.}
	\label{fig:tb_overview}
\end{figure*}

\begin{figure}
	\centering
	\begin{subfigure}{0.7\linewidth}
		\centering
		\includegraphics[width=\linewidth]{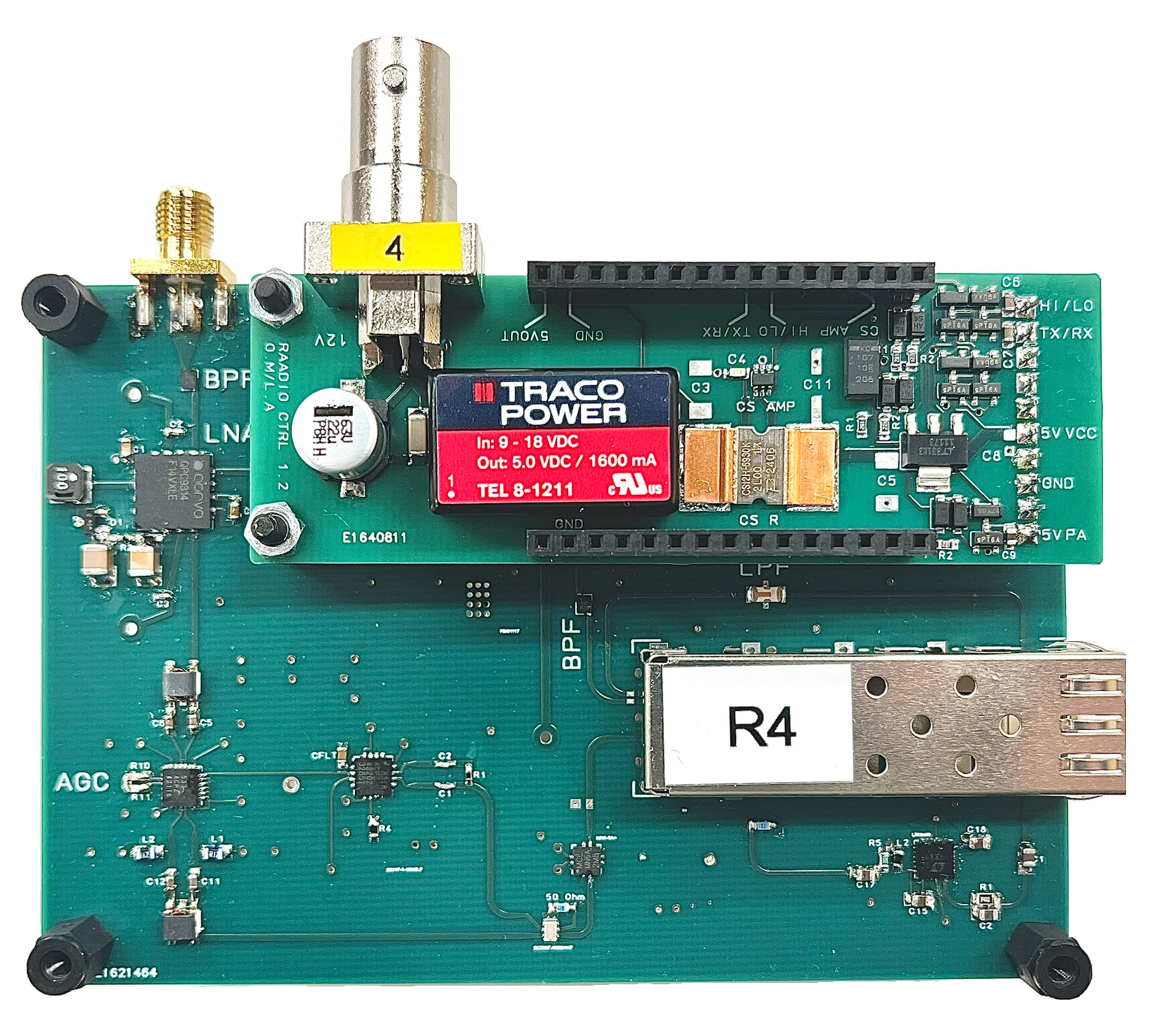}
		\caption{}
		\label{fig:rrh}
	\end{subfigure}
	\begin{subfigure}{\linewidth}
		\centering
		\includegraphics[width=\linewidth]{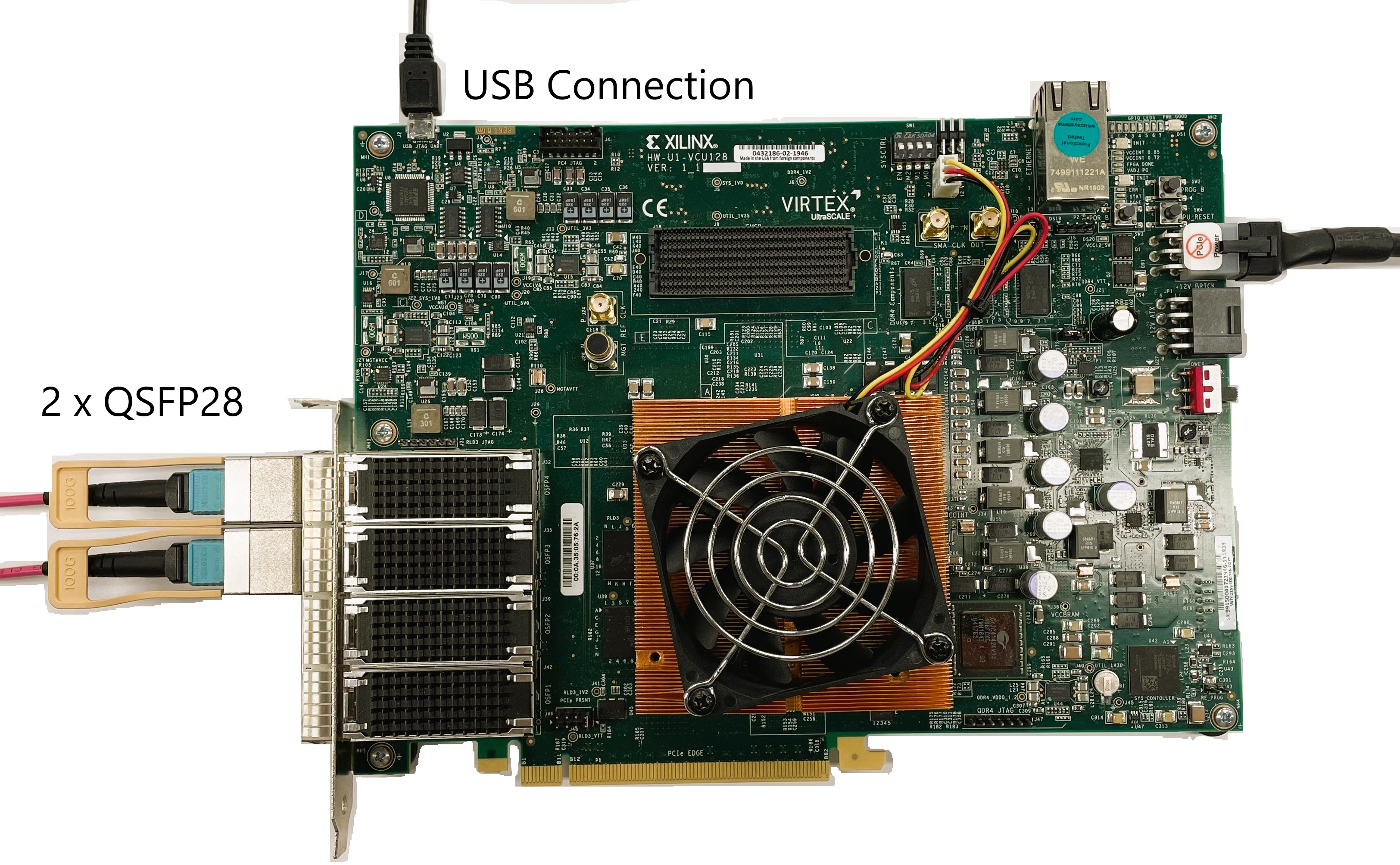}
		\caption{}
		\label{fig:cu}
	\end{subfigure}
	\caption{(a) The RRH equipped with a DC-DC converter and logic control circuit, and (b) the CU connected to QSFP28 modules.}
	\label{fig:rrh-cu}
\end{figure}

\section{The Improved $1$-Bit Radio-over-Fiber Testbed}\label{sec:testbed}
We present in this section an overview of the $1$-bit radio-over-fiber testbed used to
conduct all measurements in this paper. We then highlight the key improvements over the
testbed presented in \cite{aabel24}, which allow us to consider in this paper multi-user
transmission over larger bandwidths as well as a larger deployment scenario.

\subsection{An Overview of the Testbed}
We provide in the block diagram depicted in Fig.~\ref{fig:tb_overview} an overview of the testbed considered in this paper, which involves $6$ \glspl{rrh}
serving $2$ \glspl{ue}.
In this architecture, the \gls{cu} consists of a PC, a Xilinx FPGA, and a power supply.
Furthermore, the \glspl{rrh} are equipped with an additional circuit implementing a
\SI{12}{V} to \SI{5}{V} conversion for the DC power supply and the logic to control the
\gls{tdd} switch, the \gls{lna} gain, and the \gls{pa} (see Fig.~\ref{fig:rrh}). The PC
is responsible for all digital signal processing tasks, including signal modulation and
demodulation, dither signal generation, and \gls{sdm}.

In the uplink, \gls{rf} signals are received at the antenna port of each
\gls{rrh}.
These signals are amplified by a \gls{lna} and then fed to an \gls{agc} unit, which
ensures that the signal power remains within a prescribed range before reaching
one of the two ports of the quantizer.
Simultaneously, the \gls{cu} generates a sigma-delta modulated triangular dither signal
and sends it over the downlink optical fronthaul to the \glspl{rrh}, where it is
lowpass-filtered and fed to the other port of the quantizer.
The quantizer then generates a binary output signal, based on the sign of the difference of
the signals at its two input ports. The quantized \gls{rf} signal is then converted to the
optical domain by the \gls{sfp}28 transmitter circuitry and sent over the optical fiber
cable to the \gls{cu}.
The optical signal received at the \gls{cu} is converted by the Q\gls{sfp}28 to an
electrical signal which is sampled by the \gls{fpga} at \SI{25}{Gb/s} and written to the \gls{fpga} memory.
We use a personal computer to retrieve the memory content of the \gls{fpga} and process
the signal, which includes digital down-conversion and demodulation.
We emphasize that the \gls{agc} has a limited dynamic range,
i.e., it cannot always ensure that the \gls{rf} signal is scaled to the level required
for effective dithering.
Consequently, when the received \gls{rf} signal power falls outside the range in which the
\gls{agc} can adjust its gain, the successive dithering operation becomes ineffective
due to a suboptimal signal-to-dither power ratio.

In the downlink, the \gls{cu} performs signal modulation, digital up-conversion, and
\gls{rf} sigma-delta modulation. The 1-bit output signals from the sigma-delta modulator
are converted to the optical domain by the Q\gls{sfp}28 modules and transmitted over
individual optical fiber cables, each connecting the \gls{cu} to a single
\gls{rrh}. At each \gls{rrh}, an \gls{sfp}28 module converts the received two-level
optical signal back to the electrical domain.
The \gls{rf} signal is reconstructed by a bandpass filter and provided to a \gls{pa}.
Each \gls{rrh} is equipped with a patch antenna with \SI{6.5}{dBi} directivity.

The \gls{ue} receiver is emulated using a Keysight UXR0334A oscilloscope, while  the
\gls{ue} transmitter is emulated using two different signal generators. For single-user measurements,
we utilize the Rohde \& Schwarz
SMCV100B vector signal generator and leverage its remote control options, which facilitate
performing repeated measurements.
For multi-user measurements, we use the Agilent Technologies arbitrary waveform generator
M8190A, because it provides time-synchronized ports that enable the emulation of the
transmission from two \glspl{ue}.
Each \gls{ue} is equipped with a TE Technologies ANT-2.4-CW-HW-SMA
omnidirectional half-wave dipole antenna.

\subsection{Targeted Improvements Compared to~\cite{aabel24}}~\label{sec:key_improvements}
As shown in Table~\ref{tab:prevwork}, the measurements conducted with the testbed
described in~\cite{aabel24}\footnote{For clarity, we shall refer to it as
\textit{previous} testbed. In contrast, the testbed presented in this paper will be
referred to as \textit{upgraded} testbed.} involve signals with only $\qty{5}{MHz}$
bandwidth.
This limitation is due to the relatively low sampling rate at the \gls{cu}
($\qty{10}{Gb/s}$) in the previous testbed, which does not provide sufficiently high oversampling rate to support
larger bandwidths.
In the upgraded testbed, support for much larger bandwidths is enabled by the use of the Virtex UltraScale+ HBM VCU128 \gls{fpga}
evaluation kit, combined with the optical transceiver modules \gls{sfp}28 (supporting
a rate of \SI{25}{Gb/s}), and the Q\gls{sfp}28, which operates four parallel channels at \SI{25}{Gb/s} each (see Fig.~\ref{fig:cu}).
Note that the optimal sampling rate to achieve a certain signal quality depends on the deployment scenario, the number of RRHs and the number of active UEs. 
There is therefore not a direct translation between the sampling rate required to achieve a target \gls{evm} for a given bandwidth.

Another limitation of the previous testbed was the small deployment
area ($\qty{1}{m}\times \qty{2}{m}$).
With the upgraded testbed, we extended the area to $\qty{3.5}{m}\times \qty{4}{m}$ by introducing a new solution for distributing the power supply and the logic signals required to control the \gls{tdd} switch, the \gls{lna} gain, and the \gls{pa}.
Specifically, we equipped each \gls{rrh} with a \SI{12}{V}-to-\SI{5}{V} converter to ensure a stable \SI{5}{V} bias voltage, as well as  a transistor circuit to maintain proper timing of the logic control signals.
Both power and control signals are delivered over the same cable from a power supply located at the \gls{cu}.

\section{Single-User D-MIMO Measurements}\label{sec:su-dmimo}
One of the key advantages of D-\gls{mimo} is its ability to deliver uniform service across the coverage
area, provided that a sufficiently large number of \glspl{rrh} are deployed.
In the uplink, this is achieved because, with high probability, the signal from each \gls{ue} is
received at at a subset of \glspl{rrh} with
a sufficiently high signal-to-noise-and-interference ratio.
As already pointed out, though, in the $1$-bit radio-over-fiber architecture considered in
this paper, whenever the received \gls{rf} signal power lies outside the dynamic range of
the \gls{agc} (because it is either too low or too high), performance in terms of \gls{evm}
is drastically deteriorated because of suboptimal dithering conditions.
Hence, due to this dynamic range limitations, it is unclear whether
uniform service across the coverage area can be provided.

\begin{figure}
	\centering
	\includegraphics[width=\linewidth]{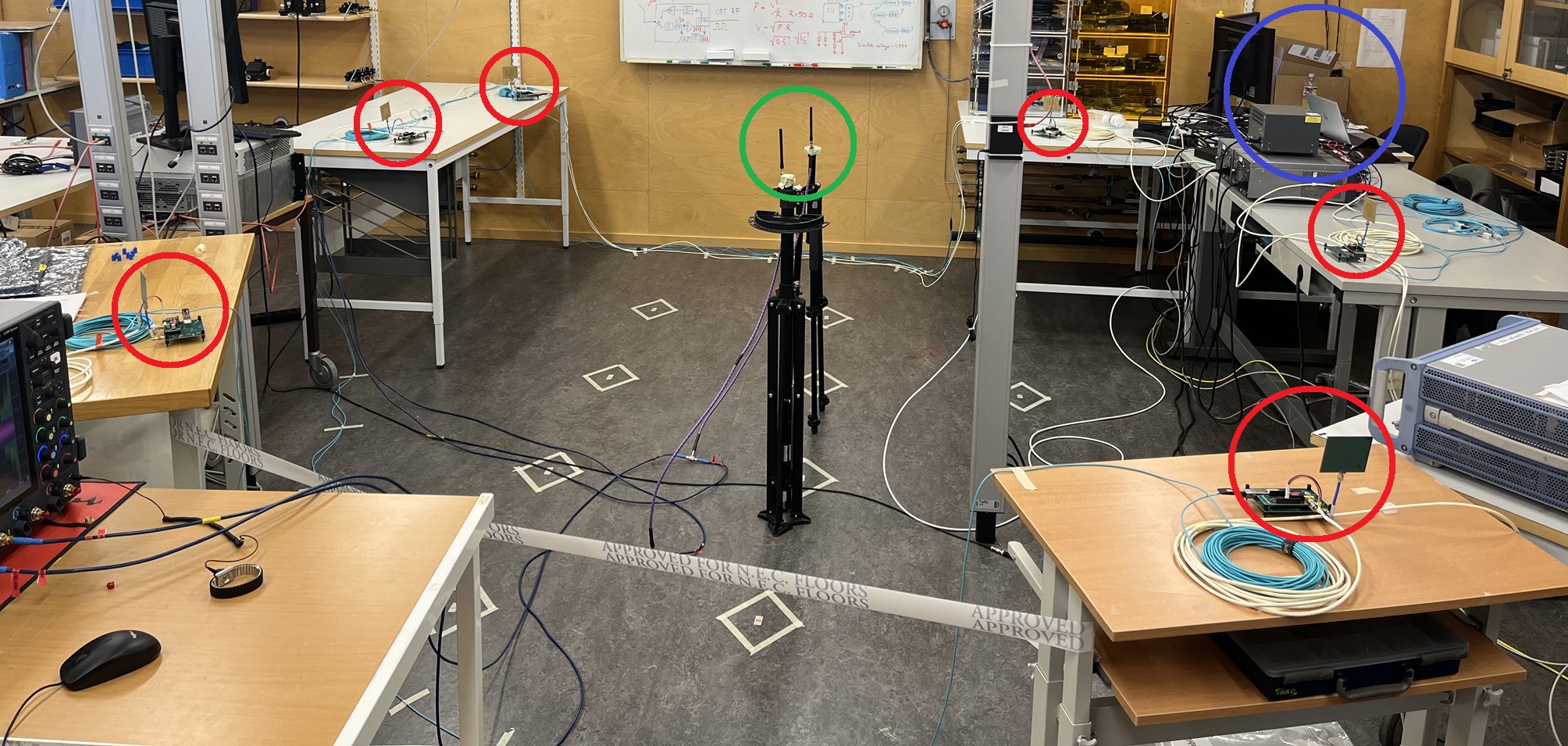}
	\caption{The deployment scenario considered in the measurement campaigns described in Section~\ref{sec:su-dmimo} and~\ref{sec:mu-dmimo}.
		The red circles mark the \glspl{rrh}, the green circle marks the \glspl{ue}, and the blue circle marks the \gls{cu}.}
	\label{fig:lab}
\end{figure}

The objective of this first investigation is therefore to use the testbed described in
Section~\ref{sec:testbed} to assess service uniformity in terms of \gls{evm}, for the case
in which the \glspl{rrh} serve a single \gls{ue}. The deployment scenario, depicted in
Fig.~\ref{fig:lab}, involves furniture, pillars, and cabinets.
This yields a rich scattering environment with both line-of-sight and non-line-of-sight
components.

\begin{figure*}
	\centering
	\begin{subfigure}{0.46\linewidth}
		\centering
		\begin{minipage}{\linewidth}
			\centering
			\includegraphics[width=\linewidth]{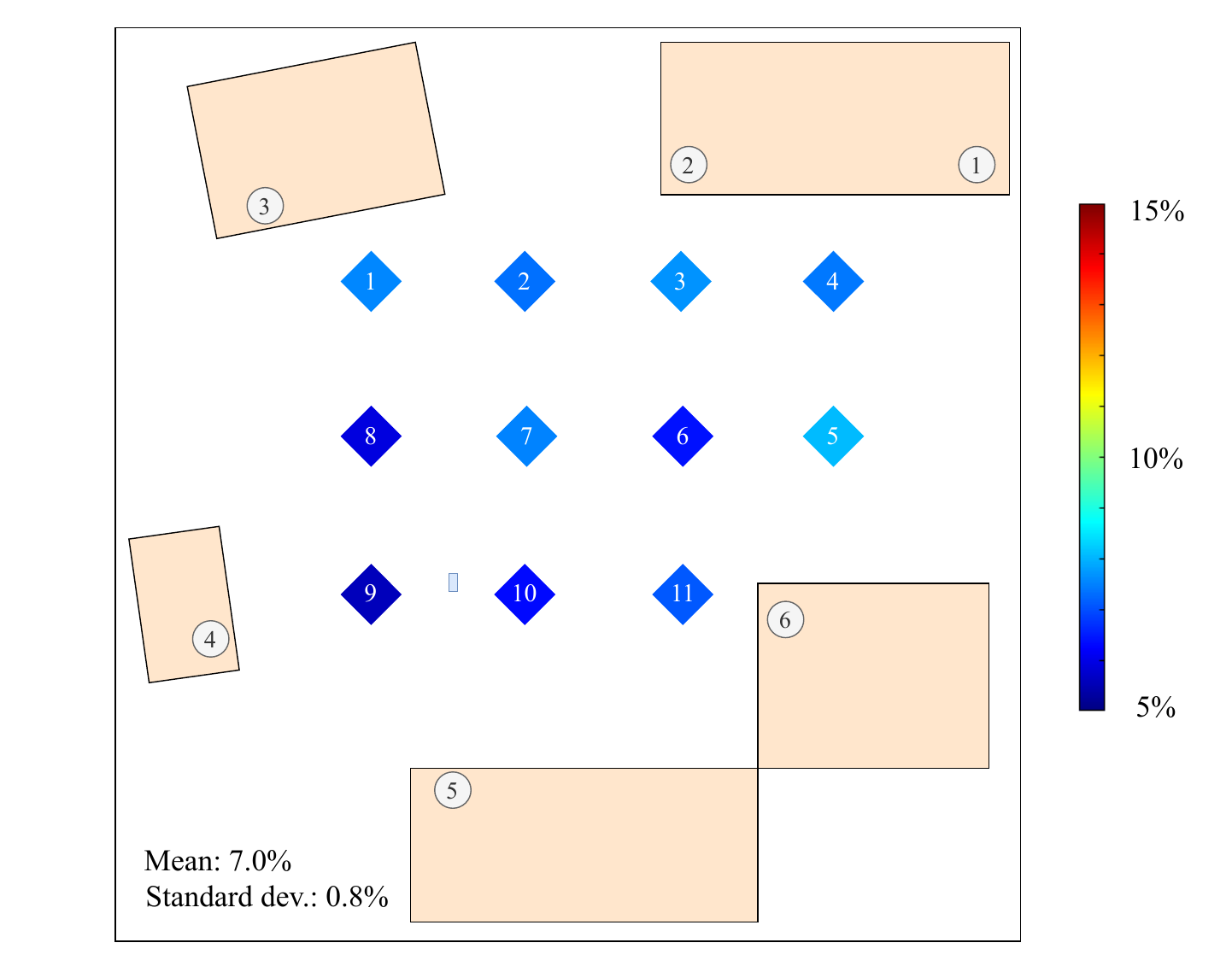}
			\caption{D-MIMO, UE TX power = \SI{0}{dBm}}
			\label{fig:dmimo-0dbm}
		\end{minipage}
		\vspace{2mm} 
		\begin{minipage}{\linewidth}
			\centering
			\includegraphics[width=\linewidth]{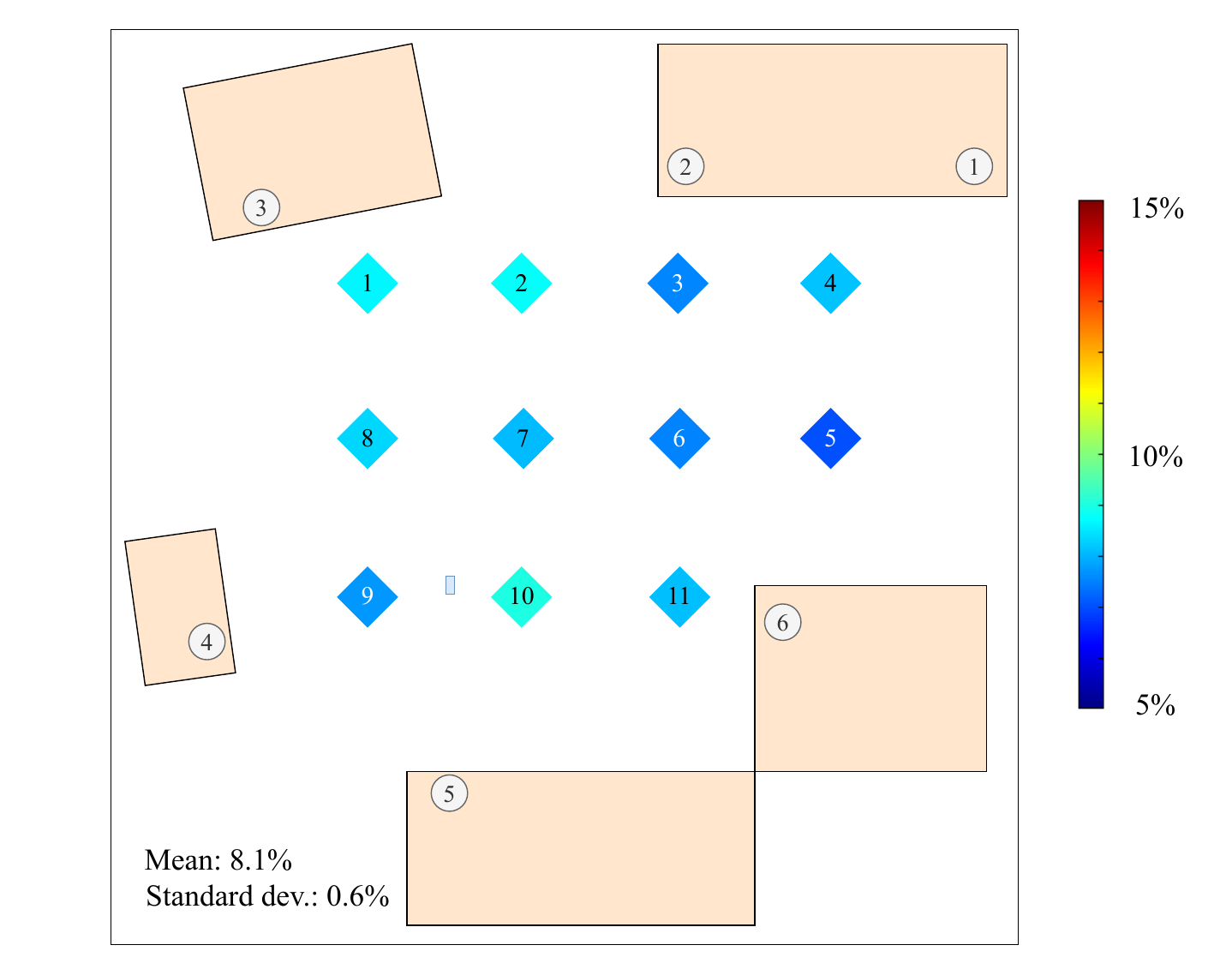}
			\caption{D-MIMO, UE TX power = \SI{-10}{dBm}}
			\label{fig:dmimo-10dbm}
		\end{minipage}
	\end{subfigure}%
	\begin{subfigure}{.45\linewidth}
		\centering
		\begin{minipage}{\linewidth}
			\centering
			\includegraphics[width=\linewidth]{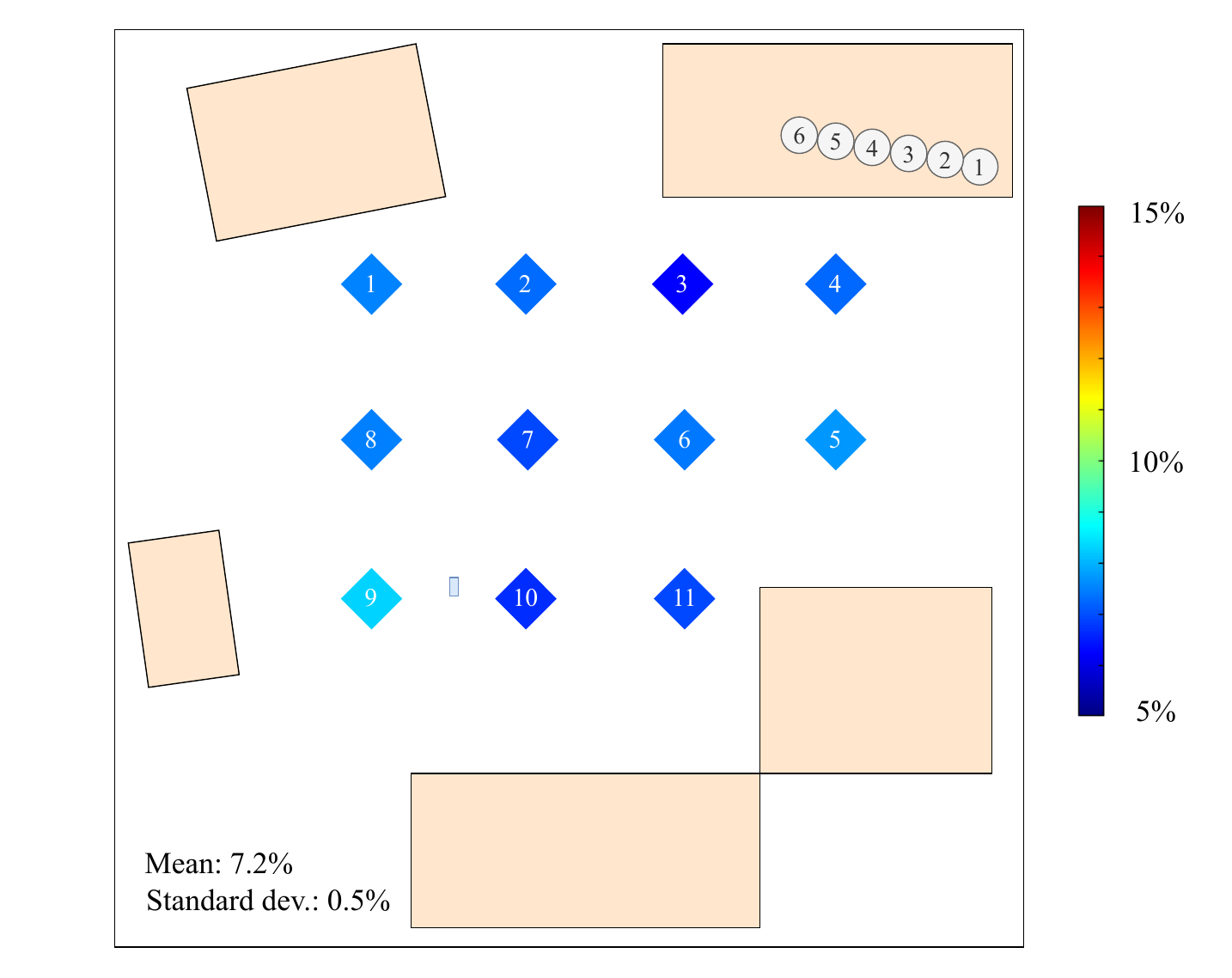}
			\caption{Co-located MIMO, UE TX power = \SI{0}{dBm}}
			\label{fig:mu-evm}
		\end{minipage}
		\vspace{2mm} 
		\begin{minipage}{\linewidth}
			\centering
			\includegraphics[width=\linewidth]{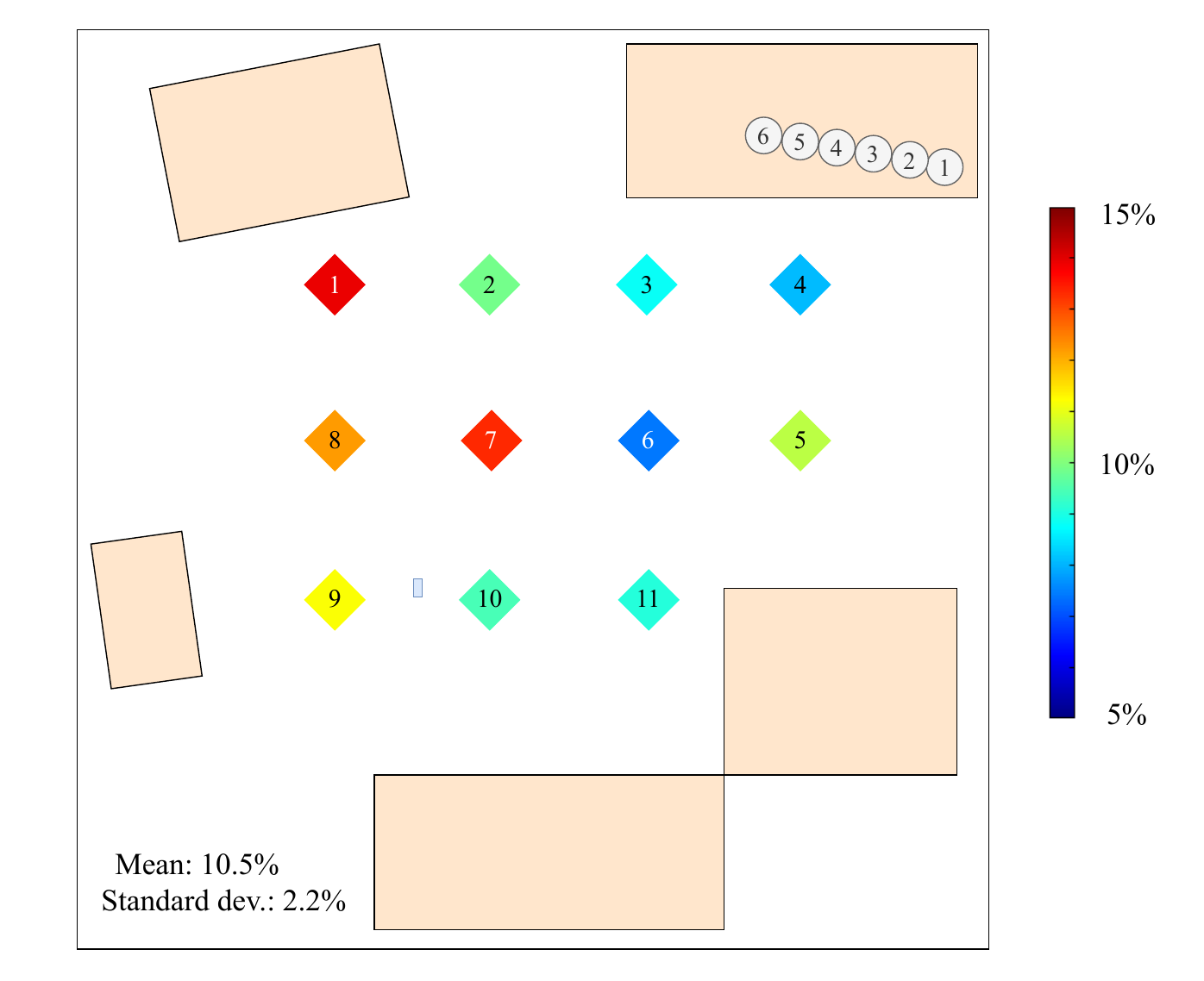}
			\caption{Co-located MIMO, UE TX power = \SI{-10}{dBm}}
			\label{fig:mu-cmimo-evm}
		\end{minipage}
	\end{subfigure}
	\caption{Each numbered square represents the measured \gls{evm} in the uplink at that position, using our $1$-bit radio-over-fiber testbed for both distributed and the co-located deployments over a $\qty{3.5}{m}\times \qty{4}{m}$ area. We consider two \gls{ue} transmit power levels: \SI{0}{dBm} and \SI{-10}{dBm}.}
	\label{fig:su-coverage-ul}
\end{figure*}

We evaluate the uplink \gls{evm} for the $11$ different
\gls{ue} positions illustrated in Fig.~\ref{fig:su-coverage-ul}, where we also depict the
positions of the \glspl{rrh}. To isolate the advantages of a distributed \gls{rrh}
deployment, we report also, for
reference, the \gls{evm} achieved with our $1$-bit radio-over-fiber testbed, for the case
in which all \glspl{rrh} are co-located over one of the tables (see
Fig.~\ref{fig:su-coverage-ul}).

The transmitted signal is an \gls{ofdm} waveform with \SI{240}{kHz} subcarrier spacing
over \SI{75}{MHz} signal bandwidth, and $16$-QAM modulation format.
The dither signal frequency is \SI{76}{MHz} and its power is \SI{-2}{dBm}.
The \gls{ue} transmits $4$ pilot symbols followed by $1$ data symbol.
The pilot symbols are used to compensate for frequency offsets and to estimate the channel
to each \gls{rrh}.
The channel estimates are then used to perform maximum ratio combining at the \gls{cu}.

\begin{figure*}
	\centering
	\begin{subfigure}{0.45\linewidth}
		\centering
		\includegraphics[width=\linewidth]{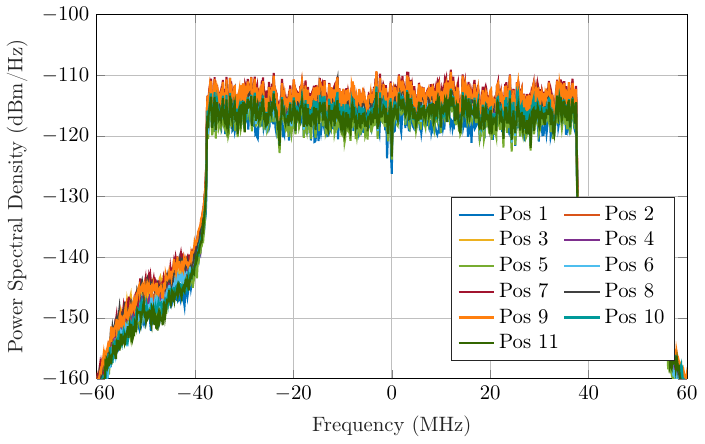}
		\caption{}
		\label{fig:su-ul-const}
	\end{subfigure}%
	\begin{subfigure}{0.45\linewidth}
		\centering
		\includegraphics[width=\linewidth]{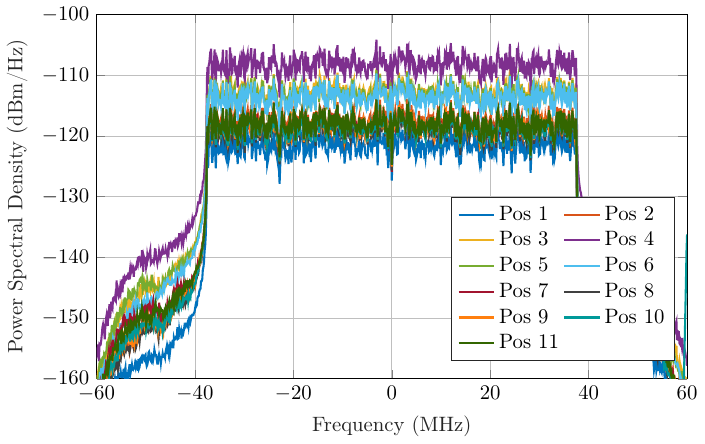}
		\caption{}
		\label{fig:su-dl-coverage}
	\end{subfigure}
	\caption{Downlink spectrum at all positions for (a) the distributed deployment and for (b) the co-located deployment.}
	\label{fig:dl-coverage}
\end{figure*}

In Fig.~\ref{fig:su-coverage-ul}, we present four heat maps of the measured \gls{evm} for
the	received uplink symbols
across all $11$ \gls{ue} positions.
We compare two \glspl{rrh} deployment strategies: distributed and co-located.
In the distributed deployment, the \glspl{rrh} are placed on tables surrounding the \gls{ue}, whereas
in the co-located deployment, all \glspl{rrh} are arranged in a row at the top-right corner of the area.
For both deployments, the \gls{ue} transmits at two power levels: $\qty{0}{dBm}$ and $\qty{-10}{dBm}$.

When the transmit power is \SI{0}{dBm}, the uplink \gls{evm}, averaged over all positions
and three consecutive measurements, is $7.0\% \pm 0.8 \%$ and $7.2\% \pm 0.5 \%$ for the
distributed and the co-located deployments, respectively. The similarity in the
\gls{evm} values and the small standard deviation across all positions
indicate that, in both the distributed and co-located deployments, the received power
remains within the dynamic range of the \gls{agc}, and an adequate
signal-to-dither ratio is maintained.
Note that a uniform quality of service can be guaranteed with our $1$-bit-over-fiber
architecture even in the co-located case, since the \gls{agc}, combined with dithering and
$1$-bit quantization, effectively nullifies the differences in path loss, provided that the
received signal power is within the \gls{agc} dynamic range.

When we lower the transmit power to $\qty{-10}{dBm}$, we observe only a small degradation of the average \gls{evm},
which is now $8.1\%\pm 0.6\%$, for the distributed deployment case. Note also that the
standard deviation remains low.
Interestingly, in the co-located case, the degradation is more significant and the
standard deviation much larger: the average \gls{evm} is $10.5\%\pm 2.2\%$.
This implies that, in the co-located scenario, an optimal signal-to-dither ratio cannot
be guaranteed at all \gls{ue} positions due to the low received power.

For completeness, we also verify that uniform quality of service can be guaranteed also in
the downlink.
We consider the setup in which $2$ downlink pilots are transmitted by each \gls{rrh} in a
round robin fashion and used at the \gls{ue} to estimate the channel.
The channel estimates are then fed back to the \gls{cu}, which uses them to perform maximum
ratio transmission.\footnote{An alternative to this approach is to perform reciprocity
calibration and downlink multi-user transmission based on uplink channel estimation.
In~\cite{aabel24}, we have demonstrated that this alternative approach is feasible with our
$1$-bit radio-over-fiber testbed.}
The transmit power at each \gls{rrh} is \SI{5}{dBm}.
We report in Fig.~\ref{fig:dl-coverage} the power spectral density of the received signal at the
\gls{ue} across all $11$ positions for both distributed and co-located deployments.
As shown in the figure, the power spectral density within the bandwidth of interest varies only by \SI{5}{dB} in the distributed deployment,
whereas the variations are more pronounced (\SI{15}{dB}) in the co-located
deployment.
This confirms that the distributed deployment results in a more uniform power
distribution across the coverage area.
Interestingly, though, the measured averaged \gls{evm}
(averaged over all positions and $3$ measurements) in the downlink is similar for both
deployments: it is $2.9\% \pm 0.3 \%$ for the distributed case, whereas it is $2.9\% \pm 0.5 \%$ for the co-located
case. The similar average \gls{evm} in both the distributed and
co-located deployments across all positions suggests that that the main cause of \gls{evm} degradation is not
the low received \gls{snr}, but rather the distortion introduced by the sigma-delta
modulator.

It is worth noting that,
for both the distributed and the co-located deployments, the
average \gls{evm} in the uplink is significantly larger than in the downlink.
This is due to two main reasons.
First, each \gls{rrh} transmits at an average power of \SI{5}{dBm}, which is higher than the
\gls{ue} transmit power.
Second, the quantization and dithering processes in the uplink
introduce more in-band quantization noise compared to the sigma-delta modulation used in
the downlink.

\section{Multi-User D-MIMO Measurements}\label{sec:mu-dmimo}
After having verified that uniform quality of service over the coverage area can be
guaranteed in the case of single-user transmission, we next extend our analysis to the
multi-user case.
Our  main aim is to determine how and in which circumstances the limitation in the dynamic range of the
\glspl{rrh} affect multi-user operations.
We consider the same distributed and co-located deployment scenarios described in
Section~\ref{sec:su-dmimo} (again, the co-located scenario is considered to disentangle the
effects caused by distributing the \glspl{rrh} from the ones caused by the signal
distortions introduced by the $1$-bit
distributed-over-fiber architecture), but assume this time that $2$ \glspl{ue} are active
within the coverage area.
We use zero-forcing precoding for downlink transmission and zero-forcing combining in
the uplink. The transmitted power from each \gls{ue} is $\SI{-5}{dBm}$.
All other parameters are as detailed in Section~\ref{sec:su-dmimo}.

\begin{figure*}
	\centering
	\begin{subfigure}{.45\linewidth}
		\centering
		\includegraphics[width=\linewidth]{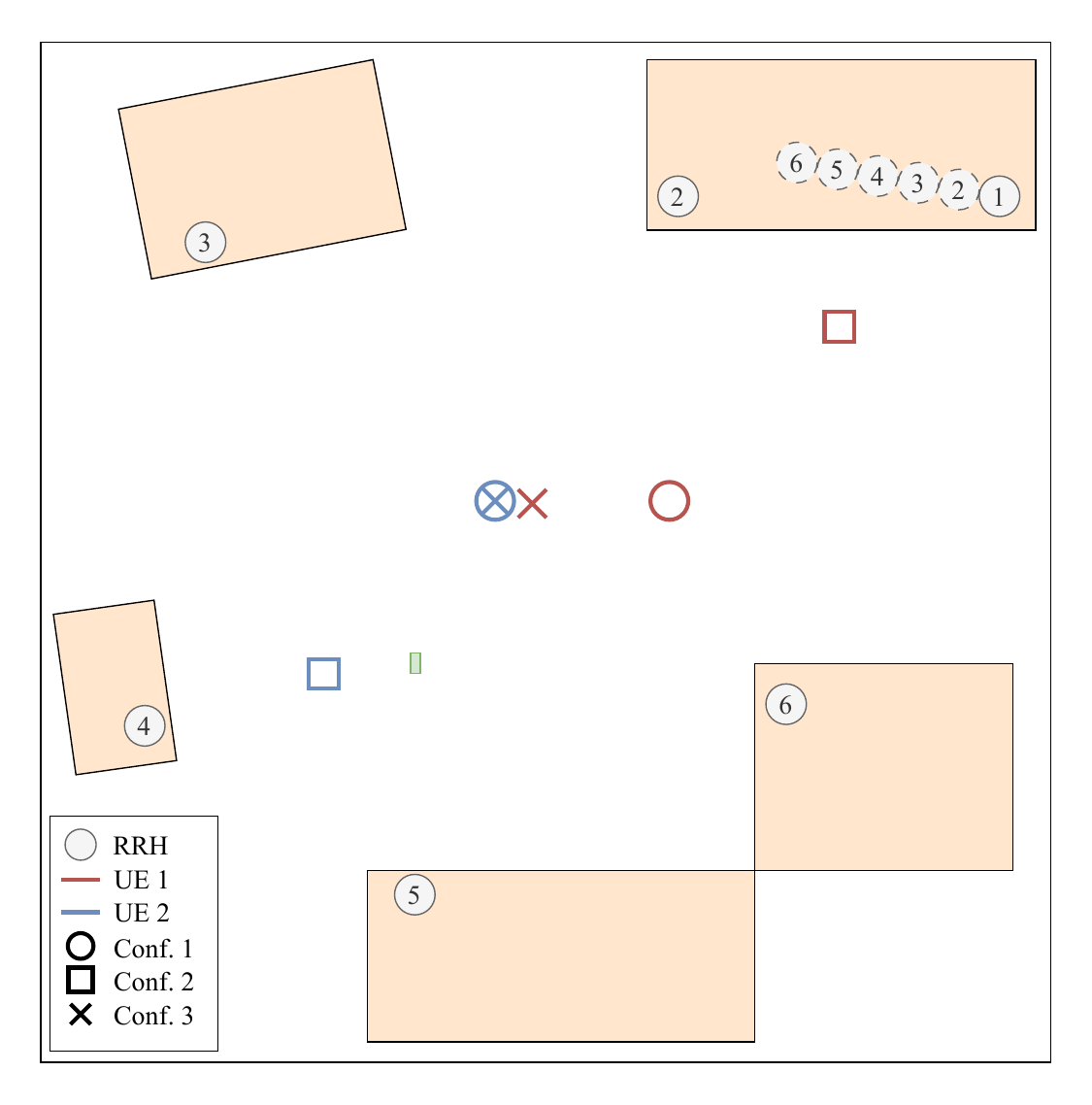}
		\caption{}
		\label{fig:mu-pos}
	\end{subfigure}%
	\begin{subfigure}{.45\linewidth}
		\centering
		\begin{minipage}{\linewidth}
			\centering
			\includegraphics[width=\linewidth]{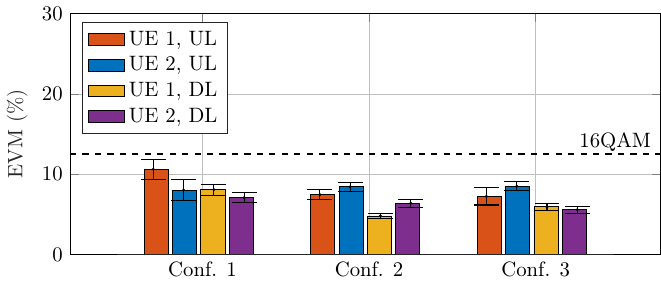}
			\caption{}
			\label{fig:mu-dmimo-evm}
		\end{minipage}
		\vspace{2mm} 
		\begin{minipage}{\linewidth}
			\centering
			\includegraphics[width=\linewidth]{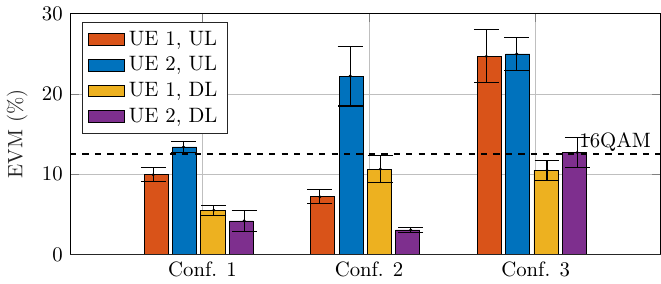}
			\caption{}
			\label{fig:mu-cmimo-evm}
		\end{minipage}
	\end{subfigure}
	\caption{(a) Positions of the two \glspl{ue} for the three configurations considered in Section~\ref{sec:mu-dmimo}; (b) average measured \gls{evm} and standard deviation for both distributed deployment and (c) co-located deployment. The dashed lines marks the 3GPP \gls{evm} requirement for 16QAM transmission.}
	\label{fig:mu}
\end{figure*}

As far as the \glspl{ue} positions are concerned, we consider the three different configurations
illustrated in Fig.~\ref{fig:mu-pos}.

\paragraph*{Configuration 1}
The \glspl{ue} are placed \SI{0.8}{m} apart in the center of the coverage area. This arrangement provides both \glspl{ue} with similar channel gains to all \glspl{rrh} in both the distributed and the co-located deployment.

\paragraph*{Configuration 2}
The \glspl{ue} are located in opposite corners of the coverage area.
In the distributed deployment, each \gls{ue} is spatially close to a different set of \glspl{rrh}.
In the co-located deployment, one \gls{ue} is significantly closer to the \glspl{rrh} than the other, leading to a significant disparity
in the received signal power.

\paragraph*{Configuration 3}
The \glspl{ue} are placed \SI{10}{cm} apart in the center of
the coverage area. This proximity increases the probability of the \glspl{ue} experiencing
highly correlated wireless channels, which makes exploiting spatial multiplexing
challenging. With this configuration, we test the capability of our $1$-bit
radio-over-fiber architecture to separate the two \glspl{ue} in the distributed and
co-located deployment cases.

Configurations 2 and 3 were chosen to analyze stress-test scenarios, in which we
expect performance degradation in at least one of the deployment scenarios,
whereas configuration 1 was chosen to explore a scenario in which it is reasonable
to expect similar performance in both deployment scenarios.

We report in Fig.~\ref{fig:mu-dmimo-evm} and  Fig.~\ref{fig:mu-cmimo-evm} the measured \gls{evm}, averaged over three consecutive measurements as well as the corresponding standard deviation, in the uplink and the downlink for both \glspl{ue}, all three configurations, and both deployment scenarios.
As shown in the figure, in the distributed deployment scenario, a relatively uniform
\gls{evm} can be achieved both in the uplink and in the downlink for all three
configurations. Furthermore, this \gls{evm} is below the level required by 3GPP for
16QAM transmission.
On the contrary, in the co-located case, we observe that the \gls{evm} changes drastically
across configurations and exceeds the 3GPP requirement in the uplink for at least one
\gls{ue}.
When comparing the \gls{evm} between the two deployments, we observe that \gls{ue} 2 experiences increased \gls{evm} in the co-located uplink scenario.
This is due to sub-optimal dithering conditions caused by the stronger power received at all \glspl{rrh} from \gls{ue} 1.
This effect is even more pronounced in Configuration 2, because the received power from \gls{ue} 2 is considerably lower.
Additionally, a notable difference in downlink performance between the two \glspl{ue} is evident in Configuration 2, indicating that the channel matrix is ill conditioned.
In Configuration 3, both the uplink and downlink \gls{evm} are large in the co-located
case, and similar for both
\glspl{ue}, suggesting that, different from the distributed case, spatial multiplexing is not possible.

To summarize, our measurements confirm that the reduced dynamic range of our
\glspl{rrh} is more deleterious in terms of performance in the multi-user case than in the
single-user case, as
illustrated by the unsatisfactory performance achievable in the co-located deployment.
Encouragingly, though, such limited dynamic range does not seem to affect significantly the performance
in the distributed deployment case.
It is also interesting to note that the difference between the \gls{evm} achieved in the
uplink and in the downlink in the multi-user case is not as significant as in the single-user case, which
suggests that the downlink performance is now limited by multi-user interference.
We expect the conclusions drawn from this measurement campaign to generalize to variations in user configurations.

\section{An Accurate Hardware Model for Our Testbed}\label{sec:tb_model}
The measurements reported in Section~\ref{sec:mu-dmimo} using our $1$-bit radio-over-fiber
testbed involve only six \glspl{rrh} and two \glspl{ue}.
In order to investigate larger scenarios, but also explore a larger range of transmit
signal power values, we present in this section an accurate hardware model for the testbed,
with the specific aim to investigate uplink performance and the impact of \gls{ue} power
control.
This model is obtained by accounting for the noise figure associated with the amplifiers, the limited dynamic range of the \gls{agc}, and the imperfections in the pulse generated by the optical transceivers.

\begin{figure*}
	\centering
	\includegraphics[width=.8\linewidth]{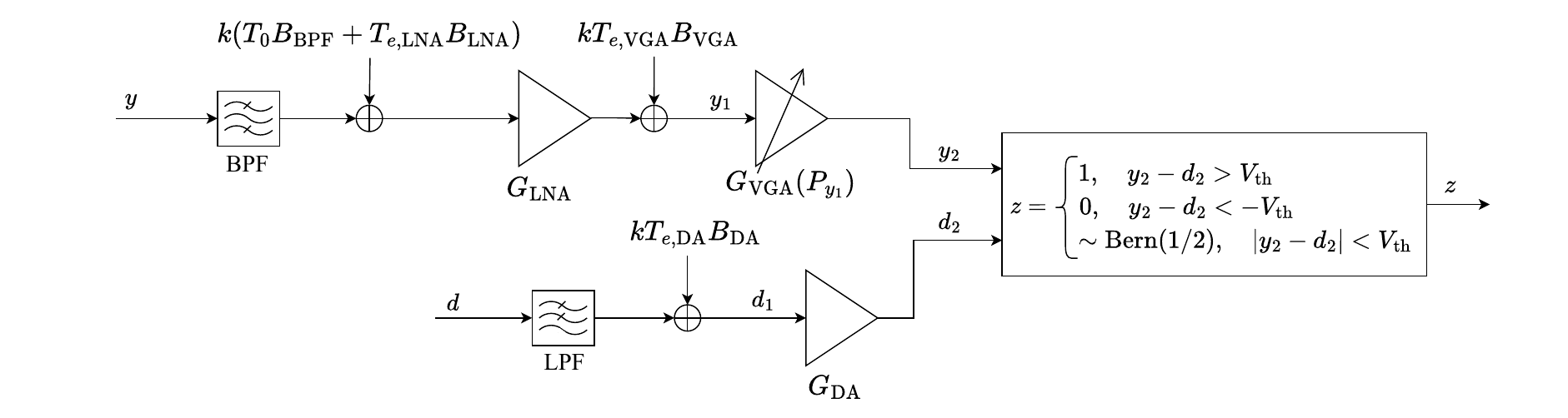}
	\caption{Block diagram of the proposed \gls{rrh} model when the \gls{rrh} operates in
		receive mode.}
	\label{fig:sim-block}
\end{figure*}

\subsection{A Model for the Uplink Fronthaul Signal} \label{sec:receiver-model}
Our hardware model is based on an accurate characterization of the signal exchanged over
the fronthaul in the uplink.
Such characterization is based on the block diagram presented in
Fig.~\ref{fig:sim-block}.
There, we describe how the inputs to each \gls{rrh}, i.e., the received
\gls{rf} signal $y$ and the sigma-delta-modulated dither signal $d$ from the \gls{cu} are processed at the \gls{rrh} to obtain the
two-level waveform $z$ at its output.
Both $y$ and $d$ pass through amplifiers, which generate noise and degrade the
\gls{snr}.
We model the power of the additive noise added by each amplifier as indicated in Fig.~\ref{fig:sim-block}, where $k$ is the Boltzmann constant, $B$ is the component-specific bandwidth and $T_e = (F-1)T_0$ is the equivalent noise temperature.
Here, $F$ denotes the noise figure~\cite[Ch. 3.5]{pozar} and $T_{0}=\qty{290}{K}$.
The values of $B$, $T_{e}$ and of the gain $G$ of each component are given in Table~\ref{tab:components}.

We next detail the modeling of two critical components: the \gls{agc} and the quantizer.

\begin{table}[]
	\centering
	\caption{Parameter values for the receiver model presented in Fig. 6.}
	\begin{tabular}{@{}c|c c c @{}}
		\toprule
		Component identifier & $B$           & $T_e$        & $G$                                  \\
		\midrule
		LNA                  & \SI{400}{MHz} & \SI{119}{K}  & \SI{24}{dB}                          \\
		VGA                  & \SI{3}{GHz}   & \SI{4867}{K} & See eq. \eqref{eq:agc} \\
		DA     & \SI{180}{MHz} & \SI{319}{K}  & \SI{15}{dB}                          \\
		LPF                  & \SI{180}{MHz} & --           & --                                   \\
		BPF                  & \SI{100}{MHz} & --           & --                                   \\
		\bottomrule
	\end{tabular}
	\label{tab:components}
\end{table}

\begin{figure}
	\centering
	\includegraphics[width=\linewidth]{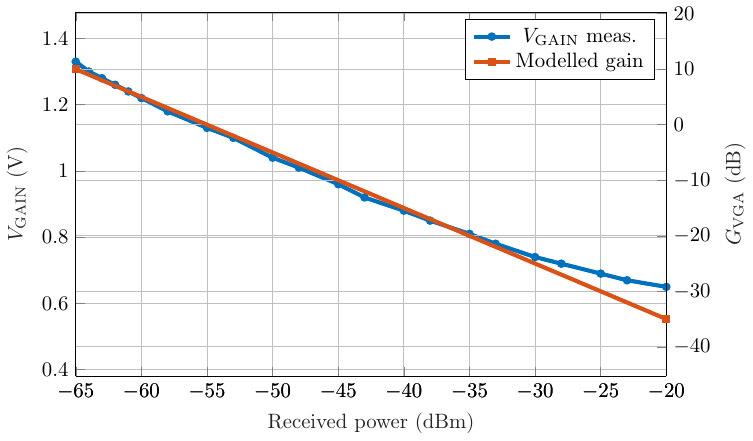}
	\caption{Measured \gls{agc} control voltage $V_{\text{GAIN}}$ and the corresponding modeled amplifier gain.}
	\label{fig:vgain}
\end{figure}

\paragraph*{\gls{agc}}\label{sec:agc}
The \gls{agc} at the \gls{rrh} consists of a \gls{vga}, from which a small portion of the
output signal is tapped and fed to a logarithmic power detector that generates the output
voltage $V_\text{GAIN}$.
This output voltage determines the gain of the \gls{vga}, which is thus a function of its input signal power.
The logarithmic nature of the power detector ensures that $V_\text{GAIN}$ is a linear
function of the input power in dB. To construct a model of the \gls{vga}, we measure its
control voltage $V_{\text{GAIN}}$ for different received power levels, measured at the
antenna port of the \gls{rrh}.
We then map this voltage to a gain value according to the component datasheet \cite{vga}.
In Fig.~\ref{fig:vgain}, we compared the measurement result with a linear model of $V_\text{GAIN}$
as a function of the received power in dB. We observe from the figure that
$V_{\text{GAIN}}$ is indeed approximately a linear function of the input power in dB, within the
dynamic range of the \gls{agc}, and that a linear model fits the measured gain control
voltage, apart from a small deviation at high received power values. According to the component
datasheet, the gain in dB of the \gls{vga} is directly proportional to $V_{\text{GAIN}}$. We
model therefore the gain of the \gls{vga} as
\begin{equation}\label{eq:agc}
	G_{\text{VGA}}\left(P_{y_1}\right) =
	\begin{cases}
		\SI{10}{dB},              & \text{if } P_{y_1} < \SI{-41}{dBm} \\
		\SI{-35}{dB},             & \text{if } P_{y_1} > \SI{4}{dBm}   \\
		- P_{y_1} - \SI{31}{dBm}, & \text{otherwise},
	\end{cases}
\end{equation}
where $P_{y_1}$ is the average power of the signal $y_1$, measured in dBm.
Note that, when the received power drives the \gls{vga} beyond its the dynamic range, the gain is modeled as a constant.

\paragraph*{Quantizer}\label{sec:qerror}
As already pointed out, we use the \gls{sfp}28 module at the \gls{rrh} to generate a
two-level pulse train, which is obtained by tracking the sign of the difference between the
filtered and
amplified \gls{rf} signal $y_{2}$ and the filtered and amplified dither signal $d_{2}$.
It is important to note that the transition between the two levels in the pulse train is not instantaneous.
Rather, it is constrained by the rise and fall time of the module.
This pulse train is converted to the electrical domain by the Q\gls{sfp}28 at the \gls{cu} and then sampled at the digital ports of the \gls{fpga}.
The distortions caused by the non-instantaneous rise and fall times imply that some of the
pulses have an amplitude that is too low to be detected by the circuitry of the
\gls{fpga}.
This results in missed pulses during the sampling operation.

To investigate this phenomenon, we compare the bit sequence sampled by
the \gls{fpga} with the one obtained by sampling the same signal using an oscilloscope and
quantizing it with the signum function. Since the oscilloscope has a high amplitude
resolution, the output of the signum function approximates well that of an ideal 1-bit
converter.
Our investigation reveals that pulses are sampled correctly by the \gls{fpga} only when the
differential input signal to the \gls{sfp}28  is larger than the threshold voltage
$V_{\text{th}}= \qty{10}{mV}$.
To model this behaviour, we assume that, whenever the magnitude of the differential input
is below $V_{\text{th}}$, the corresponding bit is drawn independently at random from a
Bernoulli distribution with parameter $1/2$ (denoted as $\text{Bern}(1/2)$):
\begin{equation}\label{eq:samp}
	z =
	\begin{cases}
		1,                    & y_2 - d_2 > V_{\text{th}}    \\
		0,                    & y_2 - d_2 < -V_{\text{th}}   \\
		\sim\text{Bern}(1/2), & |y_2 - d_2| < V_{\text{th}}.
	\end{cases}
\end{equation}
\paragraph*{Validation of the resulting model}
\begin{figure}
	\centering
	\includegraphics[width=\linewidth]{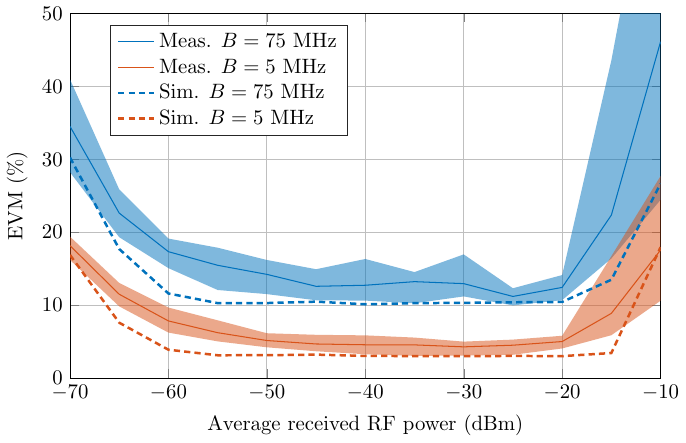}
	\caption{The measured and simulated \gls{evm} per \gls{rrh} as a function of the received power. Solid lines represent the average \gls{evm} across all \glspl{rrh}, and shaded areas represent the range between maximum and minimum \gls{evm} at each point.}
	\label{fig:dyn-range}
\end{figure}

To validate the accuracy of the proposed model, we conduct point-to-point measurements of
the \gls{evm} for the case in which a signal transmitted at different power levels is sent
from a signal generator to the antenna port of each one of the six \glspl{rrh} via a cable.
The measurement is conducted for both \SI{5}{MHz} and \SI{75}{MHz} signal bandwidth, to identify
potential bandwidth-dependent distortions.
The measured \gls{evm} is then compared to the simulated \gls{evm} for the same signal power levels.
The results are presented in Fig.~\ref{fig:dyn-range}, where dashed lines represent the simulation results, solid lines represent the average measured \gls{evm}, averaged over all $6$ \glspl{rrh}, and the shaded areas denote the range between maximum and minimum measured \gls{evm} at each point.
As shown in the figure, the simulated \gls{evm} curves closely match the minimum
measured \gls{evm} at each point for both bandwidths, indicating that the hardware model
effectively captures the noise characteristics and dynamic range of the uplink hardware in both cases.

The discrepancies in the measured \gls{evm} across the \glspl{rrh} are probably due to tuning errors when manually adjusting the output voltage from the \gls{agc}, as well as hardware irregularities caused by the soldering process.
In a few measurements at very low and very high power levels, the reported \gls{evm} is lower than the simulated value, likely due to nonlinear effects occurring when the \gls{vga} operates outside its dynamic range.

\subsection{Comparison with Infinite-Precision Architecture}
As illustrated in Fig.~\ref{fig:mu-cmimo-evm} for the colocated case, the \gls{evm}
performance of our $1$-bit radio-over-fiber architecture is unsatisfactory whenever there
is a large difference between the signal
power received by all \glspl{rrh} from each of the \gls{ue}.
In this section, we use the model for the uplink fronthaul signal presented in
Section~\ref{sec:receiver-model} to shed more light on this phenomenon.
In particular, we show that it may occur also in the distributed case, and that
the corresponding \gls{evm} degradation is particularly pronounced with our $1$-bit
radio-over-fiber architecture, whereas it would be much milder if an undistorted version
of the \gls{rf} signal at each \gls{rrh} would be made available at the
\gls{cu}.\footnote{This would be difficult to achieve, due to unavoidable distortions in
	the transmission over the optical fiber.}

\begin{figure}
	\centering
	\begin{subfigure}{\linewidth}
		\centering
		\includegraphics[width=0.66\linewidth]{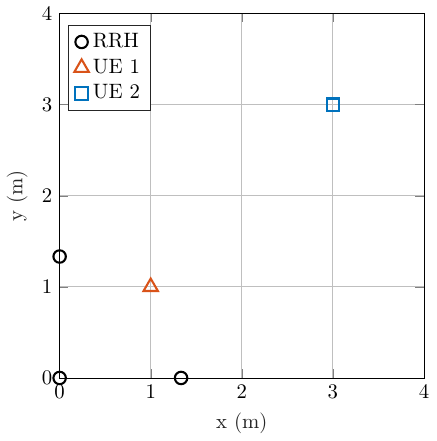}
		\caption{}
		\label{fig:2-ue-setup}
	\end{subfigure}
	\begin{subfigure}{\linewidth}
		\centering
		\includegraphics[width=\linewidth]{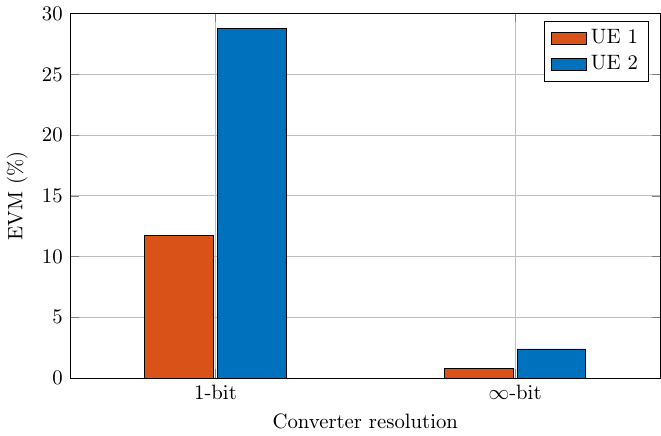}
		\caption{}
		\label{fig:2-ue-evm}
	\end{subfigure}
	\caption{(a) Simulation scenario involving two \glspl{ue} and three \glspl{rrh} (b) the \gls{evm} in the uplink achieved using 1-bit conversion and dithering or an infinite-resolution converter.}
	\label{fig:2-ue-conf}
\end{figure}

Specifically, we consider the scenario depicted in Fig.~\ref{fig:2-ue-setup}.
In this scenario, \gls{ue} 1 is in close proximity of all three \glspl{rrh}, whereas
\gls{ue} 2 is further away.
This setup ensures that the received power contribution from \gls{ue} 2 is weaker than that from \gls{ue} 1 at all \glspl{rrh}.
To model the propagation channel, we use the line-of-sight path loss model for the urban micro
scenario given in \cite{3gppChannel}. No small-scale fading is considered, for
simplicity.
The other system parameters are the same as in Section~\ref{sec:mu-dmimo}.

In Fig.~\ref{fig:2-ue-evm}, we report the \gls{evm} simulated using the model presented in
Section~\ref{sec:receiver-model}.
For reference, we also report the \gls{evm} achievable for the case in which the
undistorted \gls{rf} signal is available at the \gls{cu}, which we denote by $\infty$-bit
(infinite-resolution converters).
As expected, in both cases \gls{ue}~2 experiences a worse \gls{evm} than \gls{ue}~1;
however, while the \gls{evm} levels are still well below the 3GPP requirements in the
$\infty$-bit case, in the $1$-bit case these values are significantly exceeded by \gls{ue} 2.
This confirms that our 1-bit radio-over-fiber architecture is more sensitive to
receive-power imbalances in the uplink compared to a conventional architecture that uses
high-precision converters.

\subsection{The Benefits of \gls{ue} Power Control}
One way to address the \gls{evm} degradation illustrated in Fig.~\ref{fig:2-ue-evm}, is to implement power control.
Each \gls{rrh} sends at regular intervals beacons that are used by all \gls{ue} to estimate the
received power.
Then this information is conveyed to the \gls{cu}, that decides the power at which each
\gls{ue} should transmit (and typically, the subset of \glspl{rrh} that should serve each
\gls{ue},
according to the user-centric-design principle~\cite{cellfreebook}).
This \gls{ue} power control strategy has also the additional advantage in our $1$-bit
radio-over-fiber architecture to prevent that the \gls{agc} operates outside its dynamic
range (see Fig.~\ref{fig:dyn-range}).

\begin{figure}
	\centering
	\begin{subfigure}{\linewidth}
		\centering
		\includegraphics[width=0.66\linewidth]{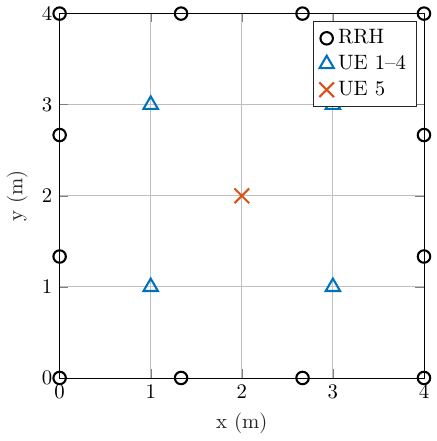}
		\caption{}
		\label{fig:5-ue-setup}
	\end{subfigure}
	\begin{subfigure}{\linewidth}
		\centering
		\includegraphics[width=\linewidth]{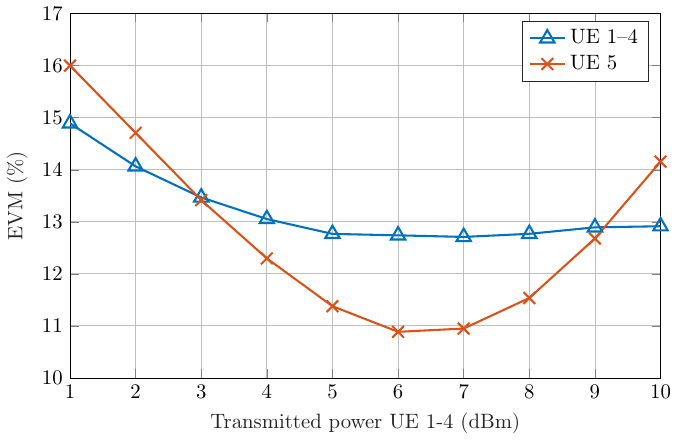}
		\caption{}
		\label{fig:5-ue-evm}
	\end{subfigure}
	\caption{(a) Simulation scenario involving $5$ \glspl{ue} and $12$ \glspl{rrh} and (b) the per-user \gls{evm} in the uplink as a function of the transmitted power from \gls{ue} 1 through 4.}
	\label{fig:2-ue-conf}
\end{figure}

To investigate the benefits of \gls{ue} power control on the per-\gls{ue} performance in
terms of \gls{evm}, we consider the
scenario depicted in Fig.~\ref{fig:5-ue-setup}, which involves $5$ \glspl{ue} and $12$
\glspl{rrh}. 
As shown in the figure, \gls{ue} 5, which is located at the center of the coverage
area, has the greatest distance to its nearest \gls{rrh} among all \glspl{ue}.
\glspl{ue} $1$--$4$ are arranged in a symmetric grid, which ensures that they experience
	similar transmission conditions.
Consequently, if all \glspl{ue} transmit at the same power level, we expect \gls{ue} 5
to exhibit the highest \gls{evm}. To confirm this intuition and explore the effect of
power control, we plot in
Fig.~\ref{fig:5-ue-evm} the simulated \gls{evm} for all \glspl{ue}, as a function
of the transmitted power of \glspl{ue} $1$--$4$, which is assumed to be identical.
In our simulation, \gls{ue}~$5$
transmits at a constant  power of \SI{10}{dBm}. The figure illustrates that when
all \glspl{ue} transmit at \SI{10}{dBm} (no power control), \gls{ue} $5$ exhibits, as expected, the highest \gls{evm}
value. As the transmitted power from \glspl{ue} $1$--$4$ is reduced, the \gls{evm} for \gls{ue}~5 improves.
In particular, all \glspl{ue} experience the same \gls{evm} when the transmit power from
\gls{ue} $1$--$4$ is about $\qty{9}{dB}$.
A further reduction of the transmit power from \glspl{ue} $1$--$4$ is initially beneficial for \gls{ue} $5$.
However, when the power falls below \qty{6}{dB}, the \gls{evm} of \gls{ue} $5$ starts to
degrade again.
This is because the channel estimates for \gls{ue} $1$--$4$ deteriorate, which impact
significantly the performance of the zero-forcing combiner.
\subsection{Towards a Power Control Algorithm}
We conclude this section by outlining some preliminary ideas for a potential power control
algorithm. A common design objective for such algorithms is to ensure
fairness among users. This is typically achieved through a max–min fairness
criterion, which seeks to maximize the minimum user spectral efficiency in the
system~\cite{cellfreebook,Ngo2017Cell-FreeCells}.
As shown in~\cite{cellfreebook}, for D-MIMO systems with infinite-precision fronthaul,
closed-form expressions for a lower bound on each users' achievable rate can be derived,
and then optimized for max-min fairness using a fixed-point algorithm.

However, these closed-form expressions do not apply to the 1-bit radio-over-fiber architecture considered in this paper,
due to the nonlinearity introduced by the 1-bit quantizer.
Although  analyzing  the spectral efficiency achievable with this architecture lies
beyond the scope of this paper, we briefly outline how such an analysis could be
conducted and the challenges it would pose for power-control design.
One possible approach, adopted in~\cite{hu24-11a} for theoretical \gls{evm}
characterization, is to linearize the quantizer
using Bussgang's theorem~\cite{bussgang52a} under the simplifying assumption that the transmitted signal
and the dither signal are Gaussian distributed.
We believe that this approach could yield a spectral efficiency lower bound that
explicitly accounts for the quantization error and for the presence of the dither signal, with the latter modelled
as a function of the received power and the \glspl{vga} gain~\eqref{eq:agc}.
The next step would be to determine whether the commonly used fixed-point optimization
algorithm can be applied to the resulting spectral-efficiency expression.

\section{Conclusion}\label{sec:concl}
We have presented hardware improvements on the $1$-bit radio-over-fiber testbed introduced
in~\cite{aabel24}, aimed at increasing both the bandwidth and the size of the deployment
scenario.
Thanks to these improvements, we were able to conduct measurement campaigns aimed at
assessing whether one can achieve the uniform quality of service promised by D-\gls{mimo}
with our $1$-bit radio-over-fiber architecture.
Our initial results are promising (see Fig.~\ref{fig:mu-dmimo-evm}).
We have also identified that uplink transmission is a potential bottleneck, due to the
limited dynamic range of the \gls{agc} (see Fig.~\ref{fig:dyn-range}), which may prevent
the $1$-bit quantizer from operating at the optimal signal-to-dither power ratio.
As illustrated in Fig.~\ref{fig:mu-cmimo-evm}, this issue yields a notable performance
degradation when there is a significant difference in the power received from the
\glspl{ue} at all \glspl{rrh}.
By leveraging the accurate simulation model introduced in Section~\ref{sec:receiver-model},
we have shown that this issue can be mitigated via \gls{ue} power control.
These findings highlight the need to  devise novel \gls{ue} power control and \gls{ue}--\gls{rrh}
association strategies that explicit account for the specific hardware limitations imposed by the
$1$-bit radio-over-fiber architecture considered in this paper.

\ifCLASSOPTIONcaptionsoff
	\newpage
\fi

\typeout{}
\bibliographystyle{IEEEtran}
\bibliography{extracted}

\vfill

\end{document}

%% file: acronyms.tex
\setabbreviationstyle[acronym]{long-short}

\glssetcategoryattribute{acronym}{glossdesc}{title}

\newacronym{cu}{CU}{central unit}
\newacronym{rrh}{RRH}{remote radio head}
\newacronym{mimo}{MIMO}{multiple-input multiple-output}
\newacronym{adc}{ADC}{analog-to-digital converter}
\newacronym{dac}{DAC}{digital-to-analog converter}
\newacronym{agc}{AGC}{automatic gain control}
\newacronym{bbu}{BBU}{baseband unit}
\newacronym{du}{DU}{distributed unit}
\newacronym{ru}{RU}{radio unit}
\newacronym{ue}{UE}{user equipment}
\newacronym{bb}{BB}{baseband}
\newacronym{rf}{RF}{radio frequency}
\newacronym{lo}{LO}{local oscillator}
\newacronym{bs}{BS}{base station}
\newacronym{evm}{EVM}{error-vector magnitude}
\newacronym{ofdm}{OFDM}{orthogonal frequency-division multiplexing}
\newacronym{snr}{SNR}{signal-to-noise ratio}
\newacronym{tdd}{TDD}{time-division duplex}
\newacronym{dc}{DC}{direct current}
\newacronym{pa}{PA}{power amplifier}
\newacronym{lna}{LNA}{low-noise amplifier}
\newacronym{sfp}{SFP}{small form-factor pluggable}
\newacronym{vga}{VGA}{variable gain amplifier}
\newacronym{if}{IF}{intermediate frequency}
\newacronym{fpga}{FPGA}{field-programmable gate array}
\newacronym{sdm}{SDM}{sigma-delta modulation}

\newglossarystyle{capitalfirst}{

	\setglossarystyle{long}

	\renewcommand*{\glossentrydesc}{\CapitalizeFirst{\glossentrydesc}}

}

